\documentclass[nopacs,preprintnumbers,amsmath,amssymb]{revtex4}
\usepackage{graphicx}
\usepackage{dcolumn}
\usepackage{bm}

\begin{document}

\begin{center}
\textbf{Solar Resonant Diffusion Waves as a Driver
of Terrestrial Climate Change}\\

Robert Ehrlich\\
George Mason University\\
Fairfax, VA 22030\\
\verb+rehrlich@gmu.edu+\\
January 4, 2006
\end{center}

\newpage

\textbf{ABSTRACT}
\\
A theory is described based on resonant thermal diffusion waves in
the sun that explains many details of the paleotemperature
record for the last 5.3 million years.  
These include the observed
periodicities, the relative strengths of
each observed cycle, and the sudden emergence in time for the 100
thousand year cycle.  Other prior work suggesting a link between
terrestrial paleoclimate and solar luminosity variations has not
provided any specific mechanism.  The particular mechanism described
here has been demonstrated empirically, although not previously
invoked in the solar context.  The theory, while not without its own
unresolved issues, also lacks most of the
problems associated with Milankovitch cycle theory.
\\

\section{paleotemperature data}

The possibility that fluctuations in solar luminosity may be
responsible for changes in global temperatures has not been
overlooked by researchers,(Lean,1997) although most explanations for
periodicities in paleotemperatures are believed to involve 
factors unrelated to solar luminosity.(Zachos,2001)  Nevertheless, some
researchers have suggested that periodic solar variability has been
the cause of global temperature cycles, with periods ranging
from the 11 year sunspot period to cycles as long as 2000
years.(Hu,2003)  Earlier reports, however, suggest no specific mechanism for long term solar periodicity.  

Paleotemperatures are inferred from the
$^{18}O/^{16}O$ ratio found in benthic sediments and
in ice cores versus depth.  In particular, it is believed that a 5 $^0C$ decrease in global temperature
corresponds to a 0.1\% increase in the $^{18}O/^{16}O$ ratio
(i.e., delta $^{18}O$) in benthic sediments.(Lisiecki,2005)  Fig. 1
shows a compilation by Lisiecki and Raymo(Lisiecki,2005) who stacked
57 sets of data on benthic sediments that
cover 5.3 million years (My) before present.  Figs. 2 and 3 show
the Fourier amplitude spectra.  
  

\section{Diffusion Waves}

Diffusion waves are an established phenomenon which has been
studied theoretically and experimentally,(Mandelis,2000)
and even for heat transport in magnetized plasmas.(Reynolds,2001)   They can occur for any process, such
as heat flow when a time-dependent source is present, in which case, the waves  represent propagating
temperature fluctuations, \(\psi(\mathbf{r},t)\), and they must satisfy the diffusion
equation.  Assuming a sinusoidal time dependence, and separating
\(\psi(\mathbf{r},t) = \phi(\mathbf{r})e^{i\omega t}\)into space
and time dependent parts, we find that \(\phi(\mathbf{r})\) satisfies:

\begin{equation}
\label{eq2}
\nabla^2 \phi  - \kappa^2 \phi = Q_0(\mathbf{r})
\end{equation}

where $Q_0(\mathbf{r})$ is the space-dependent part of a source term, and where the complex wave number $\kappa$ satisfies

\begin{equation}
\label{eq3}
\kappa = \sqrt{\frac{i\omega}{D_t}} = (1+ i)\sqrt{\frac{\omega}{2D_t}} 
\end{equation}

Note that according to eq.~\ref{eq3}, the imaginary (attenuation)
part of the complex wave number \(\kappa\) and the real (oscillatory)
part are equal, implying equality of penetration depth \(L = 2\pi
Im(\kappa)\), and wavelength \(\lambda = 2\pi /Re(\kappa)\), so
higher frequency waves are more severely damped.  For a uniform
thermal diffusivity \(D_t\)
and a spherically symmetric source 
eq.~\ref{eq2} yields the solution

\begin{equation}
\label{eq4}
\phi (r) = \frac{\sin {Re(\kappa)r}}{r}
\end{equation}

For the more general case of an isotropic medium in which
\(\kappa (r)\) is dependent only on radial distance to the origin  the
same solution applies, but \(\kappa\) must now be
replaced by its average value between the source radius $r_0$ and the r at
the wave location, i.e.,

\begin{equation}
\label{eq5}
\bar{\kappa}(r - r_0) = \int_{r_0}^{r} {\kappa (r) dr}.  
\end{equation}

\section{The solar interior}

Solar physicists use a Standard Solar Model (SSM) to calculate
many properties of the sun's interior.  According to the SSM, the
three interior regions of interest here are: (1) \emph{the core} (of
radius about 25\% of the solar radius \(r_\odot\)) where fusion occurs, (2)
the surrounding \emph{radiative zone} (extends to about \(0.7
r_\odot\)), and (3) the surrounding \emph{convection zone}, where heat
makes its way to the surface by the much more rapid process of
convective flow.  The fairly sharp boundary layer between the radiative and
convection zones is known as the ``tachocline."  

Normally, one associates diffusivity with heat flow by conduction,
whereas radiation is taken as a line-of-sight transmission process.
However, inside the sun where the mean free path for photons between
emission and subsequent reabsorption is very short (typically several
millimeters), the diffusion
description is
applicable.  Obviously, radiation is the dominant heat transfer
process inside the core and radiative zones, owing to the high temperatures
there, and the $T^4$ dependence on radiative emission and absorption rates.
The diffusivity parameter depends on a
variety of other quantities that govern the likelihood of photons
being radiated and reabsorbed.  In particular, it can be shown that
the thermal diffusivity at any radial distance as (Miesch, 2005):
$D_t = 16\sigma_{sb}T^3/(3\chi\rho^2C_P)$,
where $\sigma_{sb}$ is the Stefan-Boltzmann constant, $T$ is the
absolute temperature, $\chi$ is the opacity, $\rho$ is the density, and
$C_P$ is the specific heat.  Each of the parameters in the previous
equation can be found as a function of radial distance using data from the GONG collaboration
(Christensen-Dalsgaard,1996).

Within the SSM description, the solar core is considered to be in
a state of quasistatic equilibrium changing only on a timescale of
30 million years.  However, Grandspierre and Agoston have
noted, instabilities can be generated in a rotating plasma in the
presence of a magnetic field, and these instabilities can give rise to
thermal fluctuations, and heat waves that represent deviations from
SSM's.(Grandpierre,2005)  They further show that constraints
on the size of the core magnetic field (Friedland,2004),(Gough,1998) are such that fluctuations
can indeed arise, as a natural consequence of the
nonlinearity of the MHD equations.  

These thermal fluctuations in the form of hot
"bubbles" tend to expand, rise, and cool on timescales Grandspierre and
Agoston have estimated, which are a function of their spatial extent
and the magnitude of their temperature excess.(Grandpierre,2005)  Were the bubbles to remain small
and isolated, the diffusion waves emanating from these 
fluctuations would tend to cancel each other out.  However,
Grandspierre and Agoston have shown that the bubbles tend to merge and
grow in size, and the decay time of the waves can grow even
longer than $10^7$ years, as the perturbations smooth out to include
ever larger regions -- in effect making the solar core a region of
small amplitude thermal oscillations, instead of the quiescent state
normally assumed.(Grandpierre,2005)  

\section{Description of the theory}

We postulate small random
variations in the temperature of the outer solar core (between
$0.21r_\odot$ and $0.25 r_\odot$),(Ferro,2005) which are sufficiently smoothed out, so as
to be a function only of radial distance.  The specific noise spectrum
is unimportant, except that the noise frequencies should have slowly
changing phases over times on
a scale of $10^5 y$.  

Each of the noise frequencies may be thought of as a source
of radially symmetric diffusion waves.  When a wave
reaches the tachocline, it is reflected back to the core as an ingoing
wave. However, it should be noted that reflections in the case of diffusion waves are not of the usual
type, where well-defined wavefronts are involved, but only
resemble conventional wave reflections in the high frequency
or long path-length
limit (Mandelis,2001).  Otherwise, a considerable amount of ``phase
smearing'' prevents us from describing the reflection of discrete
wavefronts.  At a boundary, such as the tachocline, the depletion or
accumulation of photons (and the associated radially symmetric
periodic temperature fluctuation), creates an inward travelling diffusion wave.
This inward wave interferes with the next outgoing wave produced
at that frequency, and a new outward wave is generated from a
superposition of the two waves.  Again, however, the interference process is
not one of simply superimposing well-defined wavefronts, which do not
exist here.  Instead, if the inward and outward waves both
correspond to a maximum positive or negative fluctuation in photon
density (i.e., have have matched phases), the resultant temperature
fluctuation would be greatest, and hence
resonance would be achieved.  To achieve resonance the ingoing wave must
arrive back at the core in phase with the next source cycle at this
frequency, or more generally after n cycles.
Thus, the condition for resonance for the nth mode is:

\begin{equation}
\label{eq6}
Re(\bar{\kappa})2r= 2r\sqrt{\frac{\omega}{2\bar{D_t}}} = 2\pi n
\end{equation}

where n =1,2,3,..., and 2r is the round trip distance travelled by the
wave from emission to absorption back at the core.
From eq.~\ref{eq6} we see that the period $T_n$ of the nth resonant mode
\(T_n = 2\pi/\omega _n\propto 1/n^2\), so that 

\begin{equation}
\label{eq7}
T_n = T_1/n^2
\end{equation}

with the period \(T_1\) of the lowest frequency resonant mode being
twice the diffusion time \(t_d\) for photons to undergo a
random walk through the radiation zone before they reach the
tachocline.  Fortunately, \(t_d\) can be calculated from the sun's
internal density and opacity (both functions of radial
distance r) using data from the SSM.(Christensen-Dalsgaard,2005)
One the most widely-cited estimates of \(t_d\)
from Mitalas and Sills(Mitalas,1992) is
$1.7\times 10^5 y = 170 ky$.  We have repeated their calculation, and obtained 190 ky.  In
what follows, we shall use a value that brackets these two: \(t_d =180\pm 10
ky\).

Using eq.~\ref{eq7} we find the predicted periods
\(T_n\) for the nth resonant mode: 

\begin{equation}
\label{eq8}
T_n = \frac{360\pm 20}{n^2} ky
\end{equation}

Some of the periods (for n = 2, 3 and 4) given by
eq.~\ref{eq8} correspond to reported  periods in the paleotemperature data
-- see markers in fig. 2.  But before we can have confidence that resonant
solar diffusion waves may offer an explanation of periodic
variations in paleotemperatures, we must address several key
issues, including whether it is possible for the
waves to attain a large degree of resonant amplification.

\textbf{Damping of diffusion waves}  
Given that a diffusion wave
for the nth resonant mode has a phase of $2\pi n$ on returning to the
source, it will have been heavily damped, and will have an
intrinsic damping:

\begin{equation}
\label{eq9}
d_n = e^{-2\pi n}
\end{equation}

The large damping of diffusion waves might seem to make the
possibility of their playing a role in any resonant amplification
process in the sun remote.  For example, given a damping $d_n$ for a
wave after one round trip, the next outgoing wave has an amplitude
$1+d_n$, and hence the number of cycles required for the amplitude of
the nth mode to double is given as $\ln 2/d_n,$ yielding 
doubling times in ky for mode n:
$t_{\times 2} = 2n^{-2}t_d e^{2\pi n}\ln{2}$.
For $n>1$ we find doubling times  
that are longer than the age of the sun -- even if diffusion waves are
reflected 100\% from the tachocline.  Yet
the doubling times are
lengthened even more, because of damping due to ``blurry boundaries,''
since fractions of an outgoing wave are reflected back from different
parts of the tachocline, and return to different parts of the
core.  Based on the relevant
layer thicknesses, we find a phase smearing
of the nth mode: $\Delta \phi \approx \pm 2\pi n/10$.  This additional
damping, which increases with n, makes it impossible for modes $n > 4$
to survive.

%
%
%
%
%
%

\textbf{Gain due to fusion amplification.}  There is a compensating
amplification effect
that more than offsets the enormous damping, and allows resonant
diffusion waves to emerge with large amplitude in a relatively few cycles.  According to the SSM, the rate of nuclear fusion in the core of the
sun (and hence solar luminosity $L$) is proportional to $T^{4.2}$ for small
deviations from the ambient core temperature.(Bahcall,2001)  To see
the effect of this large exponent in amplifying
diffusion waves, consider an outgoing wave for mode n initially
created due to a positive core temperature fluctuation $\phi_n(r_0)$ at radius
$r_0 = 0.21 r_\odot$. In advancing a small distance $dr$ outward, the wave amplitude
is reduced by $d\phi = \kappa dr$.  However, if the
wave is absorbed by the core, it elevates the core temperature at that
r by the same amount, and hence raises the luminosity there by a
4.2-fold greater fraction, due to the $L \propto T^{4.2}$ relationship. Thus,
there is a net gain in wave amplitude over this distance $dr$ of
$3.2\kappa dr$.  The amplification factor multiplies the size of the
temperature fluctuation, and it occurs virtually instantaneously on a
diffusion time scale, so that the amplification looks like a coherent
radiation wave superimposed on a much slower moving thermal wave of
the same frequency.  As the wave proceeds outward through the 
outer core, it continues to produce an
ever increasing amount of thermal radiation, owing to the continual increase in
luminosity.  Upon reaching the edge of the core ($0.25r_\odot$) the nth
resonant mode (and the core temperature) has been amplified by a gain:
\begin{equation}
\label{eq12}
g_n = e^{3.2\bar{\kappa}\Delta r}
\end{equation}
where $\bar{\kappa}$ can be found using
 eqs.~\ref{eq3},~\ref{eq5},~\ref{eq6},
 with $r_0 = 0.21r_\odot$ and $r=0.25r_\odot$.  The same mechanism operates for the negative portion of the wave
cycle, and it amplifies reductions in core
temperature.  This
amplification mechanism is similar to the exponential
increase in intensity in the case of lasing, except that here
(1) fusion makes optical pumping unecessary, and (2) the spherical
 symmetry assures that there is no loss of waves ``out the sides,'' as occurs in a laser.  Evaluating
eq.~\ref{eq12} numerically using SSM data yields $g_n=e^{9.6n}$.

Such a large gain (more than offsetting the $e^{-2\pi n}$ damping) cannot be
taken as a realistic indicator of how large the wave can grow before
it leaves the core, but it does
suggest that a wave might be able to resonantly attain its maximum value
after only a few cycles.  It should be noted that a very similar
resonance phenomenon to that described here -- but without any
fusion gain -- has been seen empirically in
thermal-wave cavity environments.(Shen,1995)

%

%

Obviously, the theory provided here, which ignores nonthermal magnetic
interactions, and assumes an initial
temperature fluctuation that is only a function of radial distance is
highly simplified.  Far more numerous non-spherically symmetric temperature
fluctuations would not result in a similar amplification process, and
hence they will become
relatively less intense over time.  However, both the nonspherically
symmetric fluctuations and the magnetic interactions must be considered
for a full understanding of losses and gains, so as to permit a
determination of the maximum steady state radial wave amplitude.

\section{Comparison of Between Theory and Paleoclimate Data}
\label{sect4}



\textbf{Periods of each mode.}  As fig.~\ref{fig2}(markers) and table~\ref{table1} show, the
Fourier amplitude spectrum for the data shows peaks at the
theory's predicted modes n = 2,3 and 4.  
The overall data with a fifth order background subtraction to
eliminate low frequency leakage shows no indication of the n = 1 mode in fig.~\ref{fig2}.  With no background
subtraction there is a non-statistically significant suggestion of a peak corresponding to this
mode in fig.~\ref{fig3}(c) and especially (d), which
together correspond to the interval 2048 to 4095 years before present.
Were it present, a n = 1 peak (located at f = 0.28 cycles per 100 ky) could not easily be distinguished from
a steeply rising low frequency background.  

No periodicities predicted by the model are clearly absent from
the data.  It is true, however, that one small peak seen in the
overall spectrum (but not in the quartile spectra) with a
period of 19 ky (dotted arrow in fig.~\ref{fig2}) is not predicted by
the theory, and it cannot be the n =
5 mode which should be at 14.8 ky were its predicted amplitude not
zero.  Obviously, we cannot exclude the possibility that other
mechanisms, including Malinkovitch cycles, play some role in
paleotemperature periodicity.  

\textbf{Signal emergence times.}  By the signal ``emergence time'' we
refer to the amount of time required for the nth mode amplitude to
rise from the modest level of the background noise to a large enough
multiple for it to be clearly seen in the paleotemperature record.
Owing to the large gain, this amplification could occur
in a few cycles according to the theory.
We can determine the
emergence time from the data only for modes that turn on after being off for a
while.  Only one of the modes (i.e., n = 2 with a 90 to 100
ky period) clearly shows this behavior, since as seen in fig 3,
it is present only in the first and third time quartiles (solid arrows).  It is not
easy to tell from fig 1 exactly where in the vicinity of 1000
ky before the present this mode begins to appear, and just how sudden is that
appearance.

The shape of the n = 2 cycle, and the abruptness of its
emergence can
be gleaned by looking at the data for the most recent 1000 ky, after making a
subtraction of the 41 ky cycle -- see fig~\ref{fig4}.  The figure also shows
vertical error bars on the original data, and a 100 ky cycle drawn to guide the
eye.  When the data is displayed in this manner the 100 ky cycle appears
much more consistent with a sinusoidal shape than the original data for the last 1000 ky.  

It can easily be seen in fig.~\ref{fig4} that the n = 2 mode appears fairly
suddenly (between 800 and 900 ky before the present), following what appears to be a sudden phase
reversal or discontinuity.  If this discontinuity is real, the
theory has no obvious explanation for it.  However, a short emergence
time is consistent with the theory, so this
behavior offers some support.  

\textbf{Relative amplitude (and shape) of each mode.}  

We have addressed the issue of the lack of a clear signal for
the n = 1 mode, and the non-appearance of modes with $n > 4$.  Can the theory account for the relative amplitude
(and shape) of the three other modes (n = 2,3,4) that are clearly seen in
the data?  It can with the aid of two further assumptions: (1) all modes have
the same intrinsic strength (as a multiple of the background) when they are active,
and (2) only the n = 3 mode is active most of the time. 

\emph{n = 2 mode (90 ky period)}.  A look at figs.~\ref{fig3}(a) and
(c), i.e., times when both the n = 2 and 3 modes were active, shows
that the two peaks have comparable strengths. (The position of the n =
2 mode is flagged by solid and dotted arrows for those quartiles that
it appears to be ``on'' and ``off.'')  Thus, the decreased size
of this mode in the overall data is apparently
less a matter of the differences in the intrinsic strength of each
mode than what fraction of the time this mode is active. The
broadening of this peak may also be attributed to its being off
perhaps half the time.

\emph{n = 3 mode (40 ky period)}. This mode is the dominant (and
narrowest) one in the overall record because it is active at all times
-- see figs.~\ref{fig3}(a)through (d).  

\emph{n = 4 mode (22.5 ky period)}. This mode is not clearly seen in
the separate quartile spectra (figs.~\ref{fig3}(a)through (d)).
However, the fact that it appears to be split in the overall spectrum
(into two frequency peaks separated by about 7\%) has a possible
explanation that relates to a single frequency turning on and off in a
periodic manner.  For example, if we perform a Fourier Transform on a sinusoidal signal that
turns off and on say every 14 cycles, we would indeed get two separate
peaks in the frequency domain separated in frequency by 7\%. This phenomenon is not due
to a defect of the Fourier Transform, but rather that two nearby frequencies will
produce beats which mimic a single frequency turning on and
off periodically. It may simply be a coincidence, but this number of
cycles on and off for the n = 4 mode is close to what is seen for the
n = 2 mode, which appears to be alternately on and off for approximate intervals
of 1028 ky ($\approx 11$ cycles).

Finally, how to explain the smallness of the n = 4
peak compared to n = 3?   Its splitting into two separate
peaks is partly responsible, as is the fact that this mode is active
only a fraction of the time, unlike the n = 3 mode.  A third reason
for the lower n = 4 amplitude is that the damping of this mode due to
blurry boundaries (twice as serious for n = 4 than n = 3) operates
mainly as the amplification process is occurring while the wave
travels through the core, which has the effect of preventing the n = 4 mode
from reaching a large amplitude.

\textbf{Can the oscillations become large enough?}

The dominant (n = 3) mode seen in the Fourier spectrum
(fig.2) is seen to have a signal to background
ratio of about 3.4.  
This implies that the background noise at that
frequency had to be resonantly amplified by this same factor.  
Given that the complex interactions with the magnetic field that might
determine an upper limit to the size of the oscillations remains to be
worked out, it is unclear if a 3.4-fold growth (at this particular
frequency) is achievable.  This
unresolved issue does represent a current
shortcoming of the theory. 

\section{Problems with Milankovitch theory and conclusion}

In Milankovitch theory past glaciations are assumed to arise from small
quasiperiodic changes in the Earth's  orbital parameters that give
rise to corresponding changes in solar insolation, particularly in
the polar regions.  A brief discussion of five problems with this
theory are listed below, and a more detailed description of some of
them can be found
elsewhere.(Karner,2000)\\

\textbf{(a) Weak forcing problem:}  The basic problem with the theory
 s that observed
 climate variations are much more intense than the insolation changes
 can explain without postulating some very strong positive feedback
 mechanism.\\

\textbf{(b) 100 ky problem:} The preceding basic problem can be illustrated for the
case of one particular parameter -- the orbital eccentricity.  The dominant climate cycle observed during the
last million years has a roughly 100 ky period, which in Milankovitch theory is
linked to a 100 ky cycle in the eccentricity.
However, the effect of this eccentricity variation should be the
weakest of all the climate-altering changes, in view of the small
change in solar insolation it would cause.  For example, consider the Earth's
orbital eccentricity, e, which has been shown to have several periods
including one of 100 ky during which e varies in the approximate
range: $ e = 0.03 \pm 0.02.$(Quinn,1991)  The resultant solar irradiance
variation found by integrating over one orbit for each of the two extreme 
e-values is about $\pm 0.055\%$, or $\pm 0.17 w/m^2$ difference
at the top of the Earth's atmosphere.  Given that climate models show that
a one percent change in solar irradance would lead to a
$1.8^0C$ average global temperature change, then the change resulting from
a $\pm 0.055\%$ irradiance change would be a miniscule $0.1^0C$ hardly 
enough to induce a major climate event -- even with significant
positive feedback.\\

\textbf{(c) 400 ky problem:} The variations in the Earth's orbital
eccentricity show a 400 ky cycle in addition to the 100 ky cycle, with
the two cycles being of comparable strength.  Yet, the record of
Earth's climate variations only shows clear evidence for the latter.\\

\textbf{(d) Causality problem:} Based on a numerical integration of
Earth's orbit, a warming climate predates by about 10,000 years the
change in insolation than supposedly had been its cause.\\

\textbf{(e) Transition problem:} No explanation is offered for the
abrupt switch in climate periodicity from 41 ky to 100 ky that is
found to have occured about a million years ago.  Of these five
problems with Milankovitch theory, the current theory clearly shares
only (c).\\


%
In conclusion,
We have here suggested a specific mechanism involving diffusion waves
in the sun whose amplitude should grow very rapidly due to an
amplification provided by the link between solar core temperature and
luminosity.  Moreover, the phenomenon of resonant amplification of
thermal diffusion waves has been empirically demonstrated, albeit not
in the solar context.(Shen, 1995)  A number of features of the theory still remain
to be resolved, but
the theory does explain many features of the
paleotemperature record, and it appears to be free of most 
defects of the Milankovitch theory.  The theory further implicitly suggests
the existence of a new category of variable stars having extremely long
periods -- i.e., $\approx 10^4$ times longer than stellar periods 
currently considered to be ``very long.'' For some stars with $M< M_{\odot}$, their thinner radiation zones might make the predicted
periods observable.

\newpage

\section*{References}
\textbf{(Bahcall,2001)} According to the SSM, the CNO cycle contributes 1.5\%
  of the sun's luminosity, with the remainder due to the pp chain --
  see: Bahcall,J.N., Pinsonneault, M.H., and Basu, S. ``Solar Models:
current epoch and time dependences, neutrinos, and helioseismological
properties,'' Ap. J., 555, 990-1012 (2001).  At the ambient solar core temperature,
these reactions have a luminosity dependence on core temperature that
varies as of
the form $T^N$, where N is approximately 4 and 15, respectively
To find the overall dependence on temperature, we may compute
numerically the derivative $d\phi/dT$ after weighting the two reactions
by their respective abundances, giving N = 4.2.

\textbf{(Christensen-Dalsgaard,1996)} Christensen-Dalsgaard, J., The
current state of solar modeling.
Science, 272, 1286-1292 (1996).  

\textbf{(Ferro,2005)} Ferro and Lavagno have shown
that it is not the entire core of the sun that
is in a metastable state, but only its outer region, i.e.,
$r > 0.21r_\odot$.  See -- Ferro, F., Lavagno,A., Quarati,P. ``Metastable and
  stable equilibrium states of stellar electron-nuclear plasmas,''
  Phys. Lett., A336, 370-377 (2005).

\textbf{(Friedland,2004)} Friedland, A. and Gruzinov, A., ``Bounds on the
Magnetic Fields in the Radiative Zone of the Sun,'' Ap. J. 601, 570--576 (2004).

\textbf{(Gough,1998)} Gough, D.O., and MacIntyre,M.E., ``Inevitability of a magnetic field in the Sun's radiative interior,'' Nature
394, 755 (1998).

\textbf{(Grandpierre,2005)} Grandpierre, A. and Agoston, G.: ``On the onset of
  thermal metastabilities in the solar core,'' Astrophysics and
  Space Science, 298(4): 537-552 (2005).

\textbf{(Hu,2003)} Hu, F. S. et al., ``Cyclic Variation and Solar Forcing of
  Holocene Climate in the Alaskan Subarctic,'' 301, (5641), 1890 - 1893 (2003); Van Geel, B., ``The Role of Solar
  Forcing Upon Climate Change, Quaternary Sci. Rev., 18, 331--338 (1999).

\textbf{(Karner,2000)} Karner, D.B., and Muller, R.A.,``Paleoclimate:
A causality problem for Milankovitch,'' Science, 288, (5474) 2143 -
2144 (2000); BolSshakov, V.A., ``The Main contradictions and drawbacks
of the Milankovitch theory, Geophysical Research Abstracts, 5, 00721
(2003.

\textbf{(Lean,1997)}Lean, J., ``The Sun's Variable Radiation and its
Relevance for Earth,'' Ann. Rev. Astron. Astrophys., 35: 33--67 (1997).

\textbf{(Lisiecki,2005)}Lisiecki, L. E., and M. E. Raymo (2005), ``A
  Pliocene-Pleistocene stack of 57 globally distributed benthic d18O
  records,'' Paleoceanography, 20, PA1003, doi:10.1029/2004PA001071. 

\textbf{(Mandelis,2000)} Mandelis, A., ``Diffusion waves and their uses,''
  Physics Today, 53, Aug 2000, 29; Diffusion wave fields: Mathematical
  methods and Green functions, Springer-Verlag, New York, June 2001.

\textbf{(Mandelis,2001)} Mandelis, A., Nicolaides, L., and Chen, Y., ``Structure and the Reflectionless/Refractionless Nature of Parabolic Diffusion-Wave Fields,'' Phys. Rev. Lett. 87, 020801 (2001)

\textbf{(Miesch,2005)} Mark S. Miesch, ``Large-Scale Dynamics in the
  Convection Zone and Tachocline,''
\\ \verb+http://www.livingreviews.org/lrsp-2005-1+ 

\textbf{(Mitalas,1992)} Mitalas, R. and Sills, K. R., 
``On the photon diffusion time scale for the Sun,''
\emph{Astrophys. J.}, {\bf 401}, 759 -- 760 (1992).

\textbf{(Quinn,1991)} Quinn, T.R., Tremaine, S., and Duncan, M., ``A
  three million year integration of the Earth's orbit,'' Astr. J.,
  {\bf 101}, 2287 -- 2305 (1991).

\textbf{(Remhof,2003)} Remhof A., Wijngaarden R.J., and Griessen R.,
  ``Refraction and reflection of diffusion fronts,'' Phys Rev Lett. 2003 Apr
  11;90(14):145502. Epub 2003 Apr 10.

\textbf{(Reynolds,2001)} Reynolds, M. A., Morales, G. J., and Maggs, J. E.,
 ``Temperature Diffusion Waves in Plasmas,'' American Physical
  Society, 43rd Annual Meeting of the APS Division of Plasma Physics
  October 29 - November 2, 2001 Long Beach, California, abstract \#LP1.142

\textbf{(Shen,1995)} Shen, J. and Mandelis, A.
"Thermal-Wave Resonator Cavity," Rev. Sci. Instrum. 66, 4999-5005 (1995).

\textbf{(Zachos,2001)} Zachos,J., Pagani, M., Sloan,L.,
  Thomas, E. and Billups, K., ``Trends, Rhythms, and Aberrations in
  Global Climate 65 Ma to Present,'' Science, 292, (5517), 686--693 (2001).

\newpage

\begin{figure}[hb]
\includegraphics[angle=0,scale=0.75]{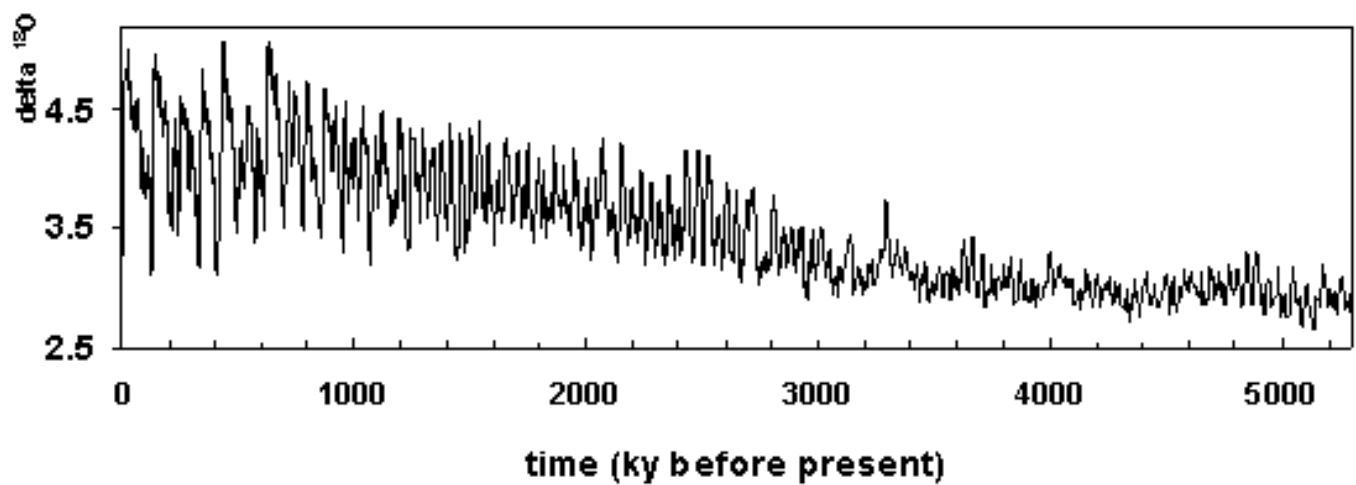}
\caption{\label{fig1} Proxy record of paleotemperatures back to 5.3 My
  before present using benthic sediment data.  An increase in the
  \(^{18}O/^{16}O\) ratio of one part per thousand
  corresponds to a $5 ^0C$
  decrease in global temperature.}
\end{figure}

\newpage

\begin{figure}[hb]
\includegraphics[angle=0,scale=0.75]{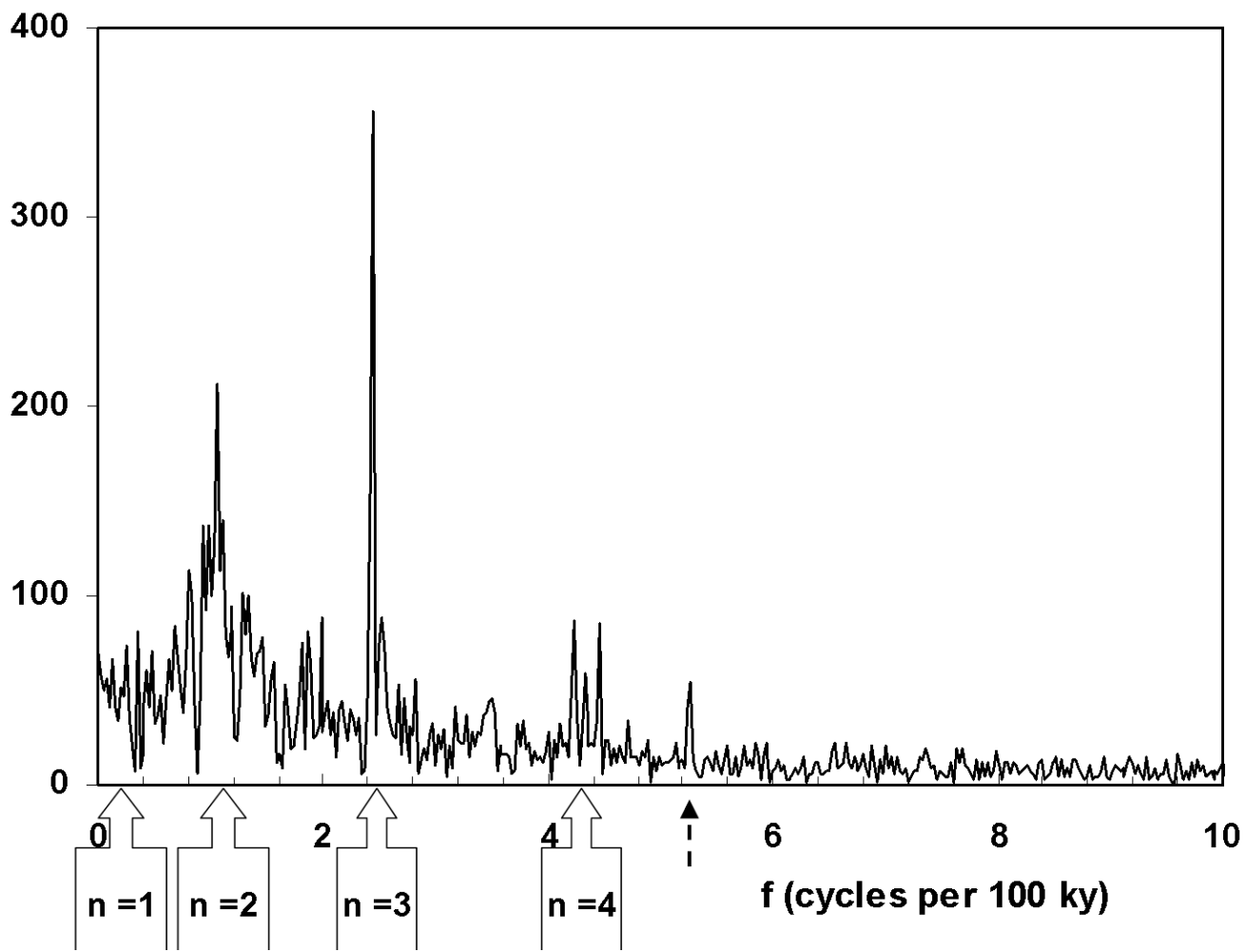}
\caption{ \label{fig2}Fourier amplitude spectrum for t = 0 to 4.1 My before
  present, after subtracting a fifth order polynomial fit to the data.
  The units for the vertical scale are arbitrary. Markers
  show our theory's predicted periodicities.}
\end{figure}

\newpage

\begin{figure}[hb]
\includegraphics[angle=0,scale=0.75]{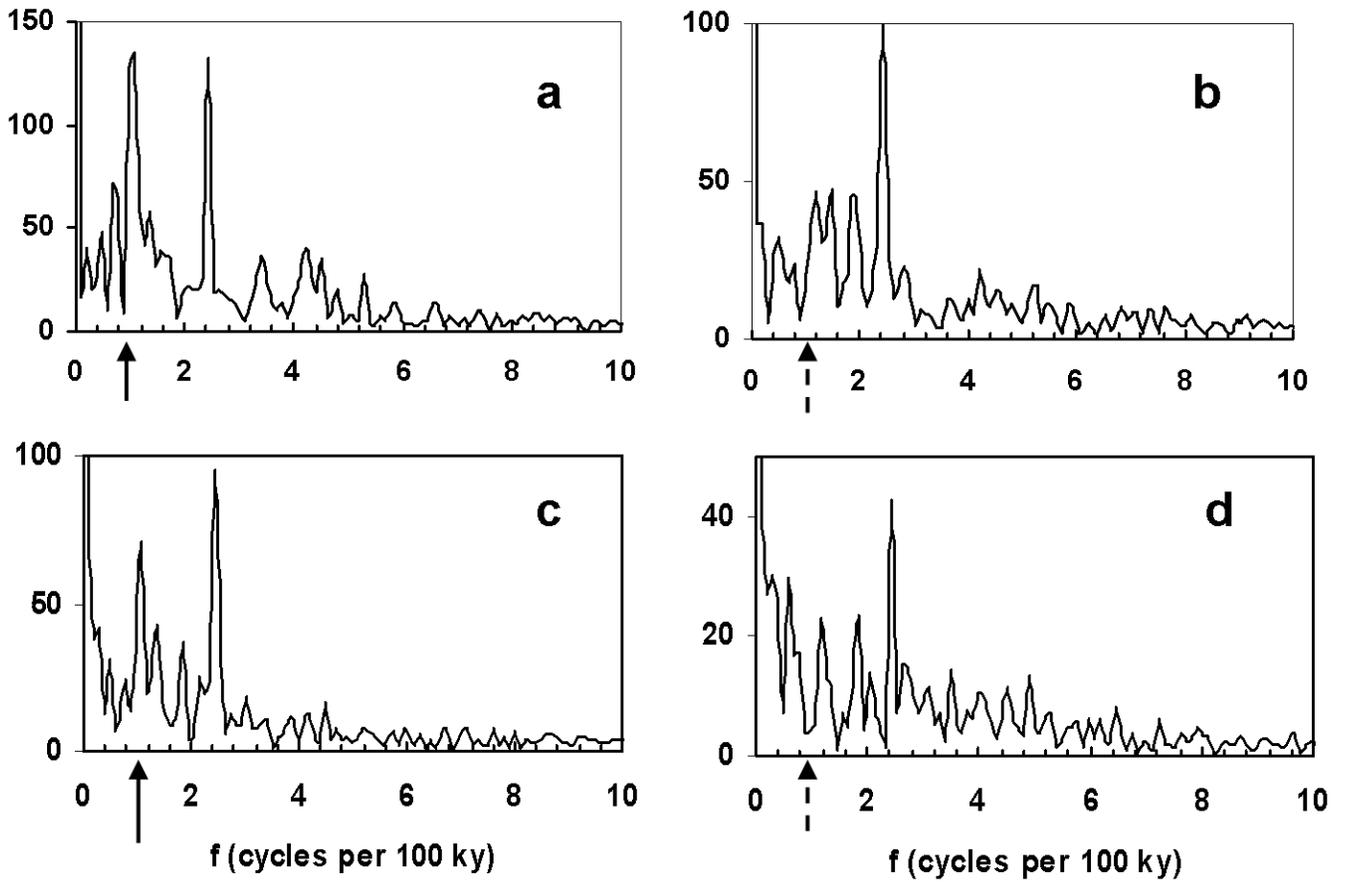}
\caption{\label{fig3} Fourier amplitude spectra for (a) 0 to 1023 ky, 
(b) 1024 to
  2047 ky, (c) 2048-3071 ky,  (d) 3072 to 4095 ky before present.  No
  background subtraction has been made.}
\end{figure}

\newpage

\begin{figure}[hb]
\includegraphics[angle=0,scale=0.70]{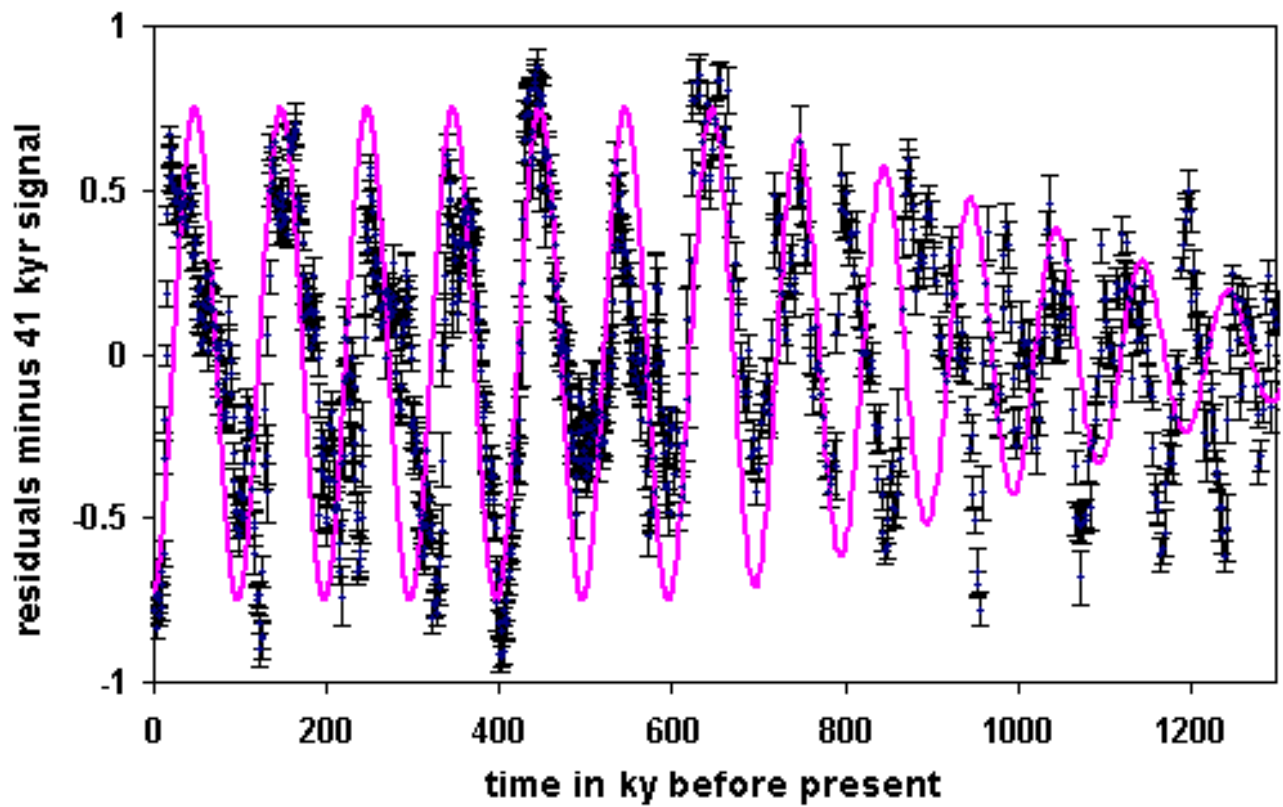}
\caption{\label{fig4} Residuals for $^{18}O$ data (including error bars) 
above a  
third order polynomial background versus time after making a subtraction 
of the 41 ky (n = 3) contribution, so as to more clearly
  observe how suddenly the n = 2 mode emerges at 800 ky.  The data
  appears to show a phase reversal at that time.}
\end{figure}

\newpage

\begin{table}
\caption{Predicted and observed periods (in ky) for modes n = 1
  to 4.  Uncertainties on the observed values are based on the half
  widths of the peaks in fig.~\ref{fig2}.\\}
\label{table1}
\begin{tabular}{|l|l|l|}
\hline
 mode n     & $T_n$ pred & $T_n$ obs\\
\hline
1&  $360\pm 20$    &???\\
2&  $90\pm 5$      &$95\pm 10$\\
3&  $40\pm 2$      &$41.0\pm 0.4$\\
4&  $22.5\pm 1.5$  &(23,7, 22.4)\\
\hline
\end{tabular}
\end{table}

\end{document}